\documentclass[12pt]{article}
\usepackage[cp1250]{inputenc}
\usepackage[verbose,a4paper,tmargin=0.5in,lmargin=1in,rmargin=1in]{geometry}
\usepackage[pdftex]{graphicx}
\usepackage{amsfonts}
\usepackage{indentfirst}
\usepackage{latexsym}
\usepackage{amsmath}
\usepackage{amsthm}
\usepackage{amssymb}
\usepackage{psfrag}
\usepackage{eufrak}
\usepackage[mathscr]{eucal} 
\usepackage{color}
\usepackage{cite}
\usepackage{verbatim}
\usepackage[all]{xy}
\usepackage{float}

\newfloat{Scheme}{tbp}{muj}

\title{Classification of the traceless Ricci tensor in 4-dimensional pseudo-Riemannian spaces of neutral signature.}

\author{$\textrm{Adam Chudecki}^{*}$}

\begin{document}

\maketitle

$*$ Center of Mathematics and Physics, Lodz University of Technology, 
\newline
$\ \ \ \ \ $ Al. Politechniki 11, 90-924 Łódź, Poland, adam.chudecki@p.lodz.pl
\newline
\newline
\newline
\textbf{Abstract}. 
\newline
The traceless Ricci tensor $C_{ab}$ in 4-dimensional pseudo-Riemannian spaces equipped with the metric of the neutral signature is analyzed. Its algebraic classification is given. This classification uses the properties of $C_{ab}$ treated as a matrix. The Petrov-Penrose types of \textsl{Plebański spinors} associated with the traceless Ricci tensor are given. Finally, the classification is compared with a similar classification in the complex case.


\section{Introduction}

The algebraic structure of Ricci tensor in general relativity was investigated by many authors (see e.g. A.Z. Petrov \cite{Petrov}, J.F. Plebański \cite{Plebanski_klasyfikacja_matter}, R. Penrose \cite{Penrose_klasyfikacja}, G.S. Hall \cite{Hall}, J.F. Plebański and J. Stachel \cite{Plebanski_Stachel}). In particular, in 1964 an algebraic classification of the traceless Ricci tensor $C_{ab}$ in real 4-dimensional Lorentzian manifolds was given by J.F. Plebański in his distinguished work \cite{Plebanski_klasyfikacja_matter} in Acta Physica Polonica. Investigation of this problem  was motivated by the obvious relation between traceless Ricci tensor and the tensor of matter $T_{ab}$. Plebański proved, that there were exactly 15 different types of the tensor of matter. In \cite{Plebanski_klasyfikacja_matter} the algebraic structure of $C_{ab}$ is investigated from several points of view. First, $C_{ab}$ is considered as a matrix. Then the structure of the so called \textsl{Plebański spinors} has been investigated. It appeared, that any tensor of matter can be represented as a superposition of three energy-momentum tensors of the electromagnetic type. Careful analysis of this fact was the third line of studies on the properties of $C_{ab}$ presented in \cite{Plebanski_klasyfikacja_matter}. 

In seventies a great deal of interest was devoted to the complex 4-dimensional spaces. It appeared, that the Plebański algebraic classification of the traceless Ricci tensor could be easily carried over to the complex spacetimes \cite{Przanowski_classification}. Since it does not make any sense to distinguish between spacelike and timelike vectors in complex spaces one could expect that the structure of $C_{ab}$ in complex spaces should be less complicated then the analogous structure in the Lorentzian case. Surprisingly, there appeared 17 different types of the traceless Ricci tensor in complex spacetime. Some of these types do not have their counterparts in real Lorentzian case. Results from \cite{Przanowski_classification} allowed one to understand better the complex relativity and the differences between complex and real manifolds. 

It is worth to note that in both papers \cite{Plebanski_klasyfikacja_matter, Przanowski_classification} the spinorial formalism has been intensively used \cite{Penrose, Plebanski_Spinors, Pleban_formalism_2}. It helped to simplify the calculations and allowed to define spinorial objects (like the Plebański spinors), which appeared to be essential in further analysis.

Recently real 4-dimensional spaces equipped with the metric of the neutral (ultrahyperbolic) signature $(++--)$ has attracted the great deal of interest. Walker and Osserman spaces, integrable systems, self-dual and anti-self-dual structures, para-Hermite and para-K\"{a}hler structures - these all concepts are related to the real 4-dimensional, neutral spaces. Especially interesting are recently discovered relation between real 4-dimensional, neutral Einstein spaces equipped with the para-K\"{a}hler structure and the 5-dimensional spaces equipped with the $(2,3,5)$-distributions \cite{Nurowski_An, Nurowski_Bor_Lamoneda}. Thus it seems that the 4-dimensional pseudo-Riemannian spaces with neutral signature will play more and more important role in theoretical physics.

Our paper is devoted to such spaces. We investigate the algebraic structure of traceless Ricci tensor $C_{ab}$ in real 4-dimensional neutral spaces. To classify $C_{ab}$ we follow the works by Plebański and Przanowski, using the same techniques. Our approach uses discrete classification (the number and type of eigenvectors of $C_{ab}$) and the continuous classification (the number and type of eigenvalues of the characteristic polynomial of $C_{ab}$). Moreover we distinguish spacelike, timelike and null eigenvectors. Plebański spinors have the same structure as a self-dual or anti-self-dual Weyl spinors and in neutral signature case they can be divided into 10 different Petrov-Penrose types. This way we obtain another criteria helpful in classification of the traceless Ricci tensor. Finally, we arrive at 33 different types of $C_{ab}$. We realize that the structure of the traceless Ricci tensor is much more rich that we could suspect.

It is well known that real spaces can be obtained from the complex spaces as the \textsl{real slices}. Real spaces with the metric of the neutral signature can be obtained from the complex ones particularly simple. It is enough to replace complex variables by real ones and holomorphic functions by real smooth ones. However, in classification of the $C_{ab}$ there appear subtle differences between complex spaces and real neutral spaces. Single complex type of $C_{ab}$ splits in a few subtypes in real case. It is related to the existence of the spacelike and timelike vectors in real spaces. In the section \ref{Classification_of_the_traceless_Ricci_tensor} we point all these differences by listing generic complex types and real types, into which these complex types split. 

We believe, that our work fills the gap left by the works of Plebański and Przanowski published in Acta Physica Polonica and will be helpful in analysis of non-Einsteinian para-Hermite and para-K\"{a}hler spaces. Some applications of ideas presented here have been already used in our previous work \cite{Chudecki_geometria_strun}.

The paper is organized as follows. In section \ref{section_Preliminaries} a portion of basic facts about the null and orthonormal tetrad in both complex and real neutral spaces is presented. Then we discuss the different types of the roots of the 4-th order polynomial and the criteria which allow to distinguish these types. The polynomials with the complex and real coefficients are both discussed. The essential difference between Petrov-Penrose classification of the 4-index, dotted and undotted totally symmetric spinors in complex and real neutral spaces are also sketched. Finally, the new symbol of the type of traceless Ricci tensor is introduced (\ref{symbol_uzywany_przeznas}). At the first glance this symbol is more complicated then the symbols used by Plebański and Przanowski in \cite{Plebanski_klasyfikacja_matter,Przanowski_classification}. We believe however, that the great number of different types of $C_{ab}$ in real neutral spaces and the complexity of the degeneration schemes (like Scheme \ref{Degeneration_scheme_of_the_Type_I}) justifies using such a symbol.

The section \ref{Classification_of_the_traceless_Ricci_tensor} is devoted to the detailed classification of the traceless Ricci tensor. We present the canonical forms of $C_{ab}$ and we discuss its possible degenerations. Also, the Petrov-Penrose type of the Plebański spinors are analyzed. The results are gathered in the tables and also the graphs of possible degenerations are presented. Concluding remarks end the paper.


\setlength\arraycolsep{2pt}
\setcounter{equation}{0}

\renewcommand{\arraystretch}{1.5}

\section{Preliminaries}
\label{section_Preliminaries}

\subsection{Formalism}

In this section we present the foundations of the formalism used in this paper. For more detailed treatment see \cite{Penrose, Pleban_formalism_2, Plebanski_Spinors}.

We consider 4-dimensional manifold $\mathcal{M}$ equipped with the metric tensor $ds^{2}$. $\mathcal{M}$ could be complex analytic differentiable manifold endowed with a holomorphic metric $ds^2$ or a real 4-dimensional smooth differentiable manifold endowed with a real smooth metric $ds^2$ of the neutral signature $(++--)$. Thus one deals with \textsl{complex relativity} ($\mathbf{CR}$) or with \textsl{real ultrahyperbolic (neutral) relativity} ($\mathbf{UR}$). 

The metric of $\mathcal{M}$ in \textsl{null tetrad} $(e^{1},e^{2},e^{3},e^{4})$ reads
\begin{equation}
\renewcommand{\arraystretch}{0.8}
ds^2 = g_{ab} \, e^{a}e^{b} = 2e^{1}e^{2} + 2e^{3}e^{4} \ , \ \ \ (g_{ab}) := 
\left(\begin{array}{cccc}
0 & 1 & 0 & 0 \\
1 & 0 & 0 & 0 \\
0 & 0 & 0 & 1 \\
0 & 0 & 1 & 0
\end{array}\right) 
\end{equation}
In \textsl{orthonormal tetrad} $(E^{1'},E^{2'},E^{3'},E^{4'})$ the metric takes the form
\begin{eqnarray}
\label{orthonormal_tetrad}
&& ds^2 =  g_{a'b'}E^{a'}E^{b'} = E^{1'}E^{1'} + E^{2'}E^{2'} - E^{3'}E^{3'} - E^{4'}E^{4'} 
\\ \nonumber
&& (g_{a'b'}) :=  \textrm{diag} (1,1,-1,-1)
\end{eqnarray}
The relation between null and orthonormal tetrad is
\begin{eqnarray}
\renewcommand{\arraystretch}{0.9}
\left\{\begin{array}{c}
\sqrt{2} \, E^{1'} = e^{1} + e^{2} \\
\sqrt{2} \, E^{2'} = e^{3} + e^{4} \\
\sqrt{2} \, E^{3'} = e^{1} - e^{2} \\
\sqrt{2} \, E^{4'} = e^{3} - e^{4} 
\end{array} \right.
\ \ \Longleftrightarrow \ \ 
\left\{\begin{array}{c}
\sqrt{2} \, e^{1} = E^{1'} + E^{3'} \\
\sqrt{2} \, e^{2} = E^{1'} - E^{3'} \\
\sqrt{2} \, e^{3} = E^{2'} + E^{4'} \\
\sqrt{2} \, e^{4} = E^{2'} - E^{4'} 
\end{array} \right.
\end{eqnarray}
In spinorial formalism the metric reads
\begin{equation}
ds^2= -\frac{1}{2} \, g_{A \dot{B}} g^{A \dot{B}}
\ \ \ \ \ \ \ \ \ \ \ \ \ A=1,2, \ \ \ \ \ \ \dot{B}= \dot{1}, \dot{2},
\end{equation}
where $g_{A \dot{B}}$ are given by
\begin{equation}
(g^{A\dot{B}}) := \sqrt{2}
\left[\begin{array}{cc}
e^4 & e^2 \\
e^1 & -e^3
\end{array}\right] 
\end{equation}
Consider now the pair of normalized undotted and dotted spinors, $(k_{A}, l_{B})$, $(k_{\dot{A}}, l_{\dot{B}})$ $k^{A}l_{A}=1$ and $k^{\dot{A}}l_{\dot{A}}=1$. They generate the new null tetrad $(\widetilde{e}^{\, 1},\widetilde{e}^{\, 2},\widetilde{e}^{\, 3},\widetilde{e}^{\, 4})$ according to the formulas
\begin{eqnarray}
 \sqrt{2} \, \widetilde{e}^{\, 1} &:=&  k_{A} l_{\dot{B}} \, g^{A \dot{B}}   \\ \nonumber
 \sqrt{2} \, \widetilde{e}^{\, 2} &:=&  l_{A} k_{\dot{B}} \, g^{A \dot{B}}   \\ \nonumber
-\sqrt{2} \, \widetilde{e}^{\, 3} &:=&  k_{A} k_{\dot{B}} \, g^{A \dot{B}}   \\ \nonumber
 \sqrt{2} \, \widetilde{e}^{\, 4} &:=&  l_{A} l_{\dot{B}} \, g^{A \dot{B}}   
\end{eqnarray}
Define the matrix $g_{aA\dot{B}}$ by the relation $g^{A\dot{B}} = g_{a}^{\ A\dot{B}}e^{a}$. The following identities hold
\begin{equation}
\label{tozsamosci_macierzy_Pauliego}
g_{aA\dot{B}}g^{bA\dot{B}} = -2 \delta_{a}^{b} \ , \ \ \ g_{aA\dot{B}}g^{aC\dot{D}} = -2 \delta_{A}^{C} \delta_{\dot{B}}^{\dot{D}}
\end{equation}
Thus we have
\begin{eqnarray}
\label{relacje_pomocne_przy_spinorach_Plebanskiego}
\renewcommand{\arraystretch}{1.1}
\left\{\begin{array}{c}
\ \ \sqrt{2} \, \widetilde{e}^{\, 1}_{\ a}  = k_{A} l_{\dot{B}} \, g_{a}^{\ A \dot{B}} \\
\ \ \sqrt{2} \, \widetilde{e}^{\, 2}_{\ a}  = l_{A} k_{\dot{B}} \, g_{a}^{\ A \dot{B}} \\
-\sqrt{2} \, \widetilde{e}^{\, 3}_{\ a} = k_{A} k_{\dot{B}} \, g_{a}^{\ A \dot{B}} \\
\ \; \sqrt{2} \, \widetilde{e}^{\, 4}_{\ a}  = l_{A} l_{\dot{B}} \, g_{a}^{\ A \dot{B}}  
\end{array} \right.
\ \ \Longleftrightarrow \ \ 
\left\{\begin{array}{c}
 \widetilde{e}_{2a} \, g^{a}_{\ A \dot{B}} = - \sqrt{2} \, k_{A} l_{\dot{B}} \\ 
\widetilde{e}_{1a} \, g^{a}_{\ A \dot{B}} = - \sqrt{2} \, l_{A} k_{\dot{B}} \\
\widetilde{e}_{4a} \, g^{a}_{\ A \dot{B}} =   \ \ \sqrt{2} \, k_{A} k_{\dot{B}} \\
 \widetilde{e}_{3a} \, g^{a}_{\ A \dot{B}} = - \sqrt{2} \, l_{A} l_{\dot{B}} 
\end{array} \right.
\end{eqnarray}

\subsection{Petrov-Penrose classification of totally symmetric 4-index spinors}
\label{subsekcja_Petrov-Penrose_classification}

Algebraic classification of totally symmetric 4-index spinors has been presented in classical paper \cite{Penrose}, see also \cite{Plebanski_klasyfikacja_matter}. It can be applied for SD (or ASD) part of the Weyl spinor $C_{ABCD}$ ($C_{\dot{A}\dot{B}\dot{C}\dot{D}}$, respectively). We use these results to classify the Plebański spinors (\ref{Plebannnn_spinorrs}). First we consider complex undotted Plebański spinors $V_{ABCD}$ and its contraction with the arbitrary 1-index spinor $\xi^{A}$: $\Omega := V_{ABCD} \xi^{A} \xi^{B} \xi^{C} \xi^{D}$. Clearly, $\Omega$ has the form $\Omega = (\xi^{2})^{4} \, \mathcal{V} (z)$ where $\mathcal{V} (z)$ is a 4-th order polynomial in $z := \xi^{1} / \xi^{2}$. Due to the fundamental theorem of algebra $\Omega$ can be always brought to the factorized form $\Omega = (\alpha_{A} \xi^{A}) (\beta_{B} \xi^{B})(\gamma_{C} \xi^{C})(\delta_{D} \xi^{D})$. Because of the arbitrariness of $\xi^{A}$ we find
\begin{equation}
V_{ABCD} = \alpha_{(A} \beta_{B} \gamma_{C} \delta_{D)}
\end{equation}
In general 1-index spinors $\alpha_{A}$, $\beta_{A}$, $\gamma_{A}$ and $\delta_{A}$ are mutually linearly independent. Such case corresponds to the case when the polynomial $\mathcal{V}(z)$ has four different roots. The possible coincidences between spinors $\alpha_{A}$, $\beta_{A}$, $\gamma_{A}$ and $\delta_{A}$ brought us to the well-known \textsl{Petrov-Penrose diagram} (Scheme \ref{Petrov_Penrose_diagram}).
\begin{Scheme}[!h]
\begin{displaymath}
\xymatrixcolsep{0.8cm}
\xymatrixrowsep{0.8cm}
\xymatrix{
\alpha_{(A} \beta_{B} \gamma_{C} \delta_{D)}   \ar[d] \ar[dr] & & 
\\
\alpha_{(A} \alpha_{B} \beta_{C} \beta_{D)}  \ar[d] \ar[dr] & \alpha_{(A} \beta_{B} \gamma_{C} \gamma_{D)} \ar[l] \ar[d] \ar[dr]  
\\
0 & \alpha_{A} \alpha_{B} \alpha_{C} \alpha_{D} \ar[l] & \alpha_{(A} \beta_{B} \beta_{C} \beta_{D)} \ar[l]
}
\end{displaymath} 
\caption{Petrov-Penrose diagram.}
\label{Petrov_Penrose_diagram}
\end{Scheme}

In complex case there are 6 different Petrov-Penrose types of the spinor $V_{ABCD}$. On the other hand, if we consider real totally symmetric 4-index spinor $V_{ABCD}$, then the scheme of the roots of $\mathcal{V}(z)$ is more complicated. There appear 10 different Petrov-Penrose types. The symbols which are usually used as a abbreviations of the corresponding Petrov-Penrose types of spinor $V_{ABCD}$ and the scheme of the roots of the polynomial $\mathcal{V}(z)$ are gathered in the Table \ref{typy_V_ABCD}. (In the Table \ref{typy_V_ABCD} $Z$ means that the root is complex while $R$ stands for the real root, the power denotes the multiplicity of corresponding root, spinors $\alpha_{A}$, $\beta_{A}$, $\gamma_{A}$ and $\delta_{A}$ are complex, spinors $\mu_{A}$, $\nu_{A}$, $\xi_{A}$ and $\zeta_{A}$ are real, bar stands for the complex conjugation).
\begin{table}[!h]
\begin{center}
\begin{tabular}{|c|c|c|c|c|c|}   \hline
\multicolumn{3}{|c|}{Complex case}  & \multicolumn{3}{|c|}{Real case}  \\  \hline
Type & $V_{ABCD}=$  &  Roots of $\mathcal{V} (z)$ & Type & $V_{ABCD}=$  & Roots of $\mathcal{V} (z)$ \\ \hline
$[\textrm{I}]$    & $\alpha_{(A} \beta_{B} \gamma_{C} \delta_{D)}$  & $Z_{1}Z_{2}Z_{3}Z_{4}$ &
$[\textrm{I}]_{r}$   & $\mu_{(A} \nu_{B} \xi_{C} \zeta_{D)}$  &  $R_{1}R_{2}R_{3}R_{4}$ \\ \cline{4-6}
& & &
$[\textrm{I}]_{rc}$  & $\mu_{(A} \nu_{B} \alpha_{C} \bar{\alpha}_{D)}$  &  $R_{1}R_{2}Z\bar{Z}$ \\ \cline{4-6}
& & &
$[\textrm{I}]_{c}$   & $\alpha_{(A} \bar{\alpha}_{B} \beta_{C} \bar{\beta}_{D)}$ &  $Z_{1}\bar{Z}_{1}Z_{2}\bar{Z}_{2}$ \\ \hline
$[\textrm{II}]$   & $\alpha_{(A} \beta_{B} \gamma_{C} \gamma_{D)}$  & $Z_{1}Z_{2}Z_{3}^{2}$ &
$[\textrm{II}]_{r}$  & $\mu_{(A} \nu_{B} \xi_{C} \xi_{D)}$  &  $R_{1}R_{2}R_{3}^{2}$ \\ \cline{4-6}
& & &
$[\textrm{II}]_{rc}$ & $\mu_{(A} \mu_{B} \alpha_{C} \bar{\alpha}_{D)}$ &  $R^{2}Z\bar{Z}$ \\ \hline
$[\textrm{D}]$    & $\alpha_{(A} \alpha_{B} \beta_{C} \beta_{D)}$   & $Z_{1}^{2}Z_{2}^{2}$ &
$[\textrm{D}]_{r}$   & $\mu_{(A} \mu_{B} \nu_{C} \nu_{D)}$   &  $R_{1}^{2} R_{2}^{2}$ \\ \cline{4-6}
& & &
$[\textrm{D}]_{c}$   & $\alpha_{(A} \alpha_{B} \bar{\alpha}_{C} \bar{\alpha}_{D)}$ &  $Z^{2} \bar{Z}^{2}$ \\ \hline
$[\textrm{III}]$  & $\alpha_{(A} \beta_{B} \beta_{C} \beta_{D)}$    & $Z_{1}Z_{2}^{3}$ &
$[\textrm{III}]_{r}$ & $\mu_{(A} \nu_{B} \nu_{C} \nu_{D)}$    &  $R_{1}R_{2}^{3}$  \\ \hline
$[\textrm{N}]$    & $\alpha_{A} \alpha_{B} \alpha_{C} \alpha_{D}$   & $Z^{4}  $ &
$[\textrm{N}]_{r}$   & $\mu_{A} \mu_{B} \mu_{C} \mu_{D}$   &  $R^{4}  $ \\ \hline
$[-]$                & $0$                                             &  $-$ &
$[-] $            & $0$                                             & $-$ \\ \hline
\end{tabular}
\caption{Petrov-Penrose types of complex or real totally symmetric 4-index spinor.}
\label{typy_V_ABCD}
\end{center}
\end{table}

It is clear that the Petrov-Penrose types of both real and complex totally symmetric 4-index spinors $V_{ABCD}$ are related to the nature of roots of the corresponding polynomial $\mathcal{V} (z)$. It is well known that such a 4-th order polynomial can be always brought to the \textsl{canonical form}. The criteria which allow to distinguish the scheme of roots of the 4-th order polynomial in the canonical form are discussed in the next subsection.

Of course, similar classification can be applied for the dotted 4-index spinors $V_{\dot{A}\dot{B}\dot{C}\dot{D}}$ and for the "dotted" polynomial $\dot{\mathcal{V}} (\dot{z})$.

\subsection{Traceless Ricci tensor}

The relation between traceless Ricci tensor $C_{a b}$ and its spinorial image $C_{AB\dot{C}\dot{D}}$ reads
\begin{equation}
C_{ab} = g_{a}^{\ A\dot{C}} g_{b}^{\ B\dot{D}} \, C_{AB\dot{C}\dot{D}} \ \Longleftrightarrow \ C_{AB\dot{C}\dot{D}} = \frac{1}{4} C_{ab} \, g^{a}_{\ A\dot{C}} g^{b}_{\ B\dot{D}}
\end{equation}
(compare (\ref{tozsamosci_macierzy_Pauliego})). Using spinorial image $C_{AB\dot{C}\dot{D}}$ of the traceless Ricci tensor $C_{ab}$ one defines the undotted and dotted \textsl{Plebański spinors} by the relations \cite{Plebanski_Spinors, Plebanski_klasyfikacja_matter}
\begin{equation}
\label{Plebannnn_spinorrs}
V_{ABCD} := 4 \, C_{(AB}^{\ \ \ \ \; \dot{M}\dot{N}} C_{AC)\dot{M}\dot{N}} \ , \ \ \ 
V_{\dot{A}\dot{B}\dot{C}\dot{D}} = 4 \, C_{MN(\dot{A}\dot{B}} C^{MN}_{\ \ \ \, \dot{C}\dot{D})}
\end{equation}
Dotted and undotted Plebański spinors are totally symmetric $V_{ABCD} =V_{(ABCD)}$ and $V_{\dot{A}\dot{B}\dot{C}\dot{D}} = V_{(\dot{A}\dot{B}\dot{C}\dot{D})}$. 

The characteristic polynomial of the matrix $C^{a}_{\ b}$ of the traceless Ricci tensor reads
\begin{equation}
 \mathcal{W} (x) := \det (C^{a}_{\ b} - x \delta^{a}_{\ b}) = \sum_{i=0}^{4} (-1)^{i} \, \underset{[i]}{\mathbb{C}} \, x^{4-i} \equiv \underset{[0]}{\mathbb{C}}x^4 - \underset{[1]}{\mathbb{C}} x^3+ \underset{[2]}{\mathbb{C}} x^2 - \underset{[3]}{\mathbb{C}} x + \underset{[4]}{\mathbb{C}}
\end{equation}
where the coefficients $\underset{[i]}{\mathbb{C}}$ are given by
\begin{equation}
\underset{[0]}{\mathbb{C}} := 1 \ , \ \ \ \underset{[k]}{\mathbb{C}} := C^{a_{1}}_{\ \; [a_{1}} ... C^{a_{k}}_{\ \; a_{k}]} \ , \ \ \ k=1,2,3,4
\end{equation}
Since the matrix $C^{a}_{\ b}$ is traceless we find that $\underset{[1]}{\mathbb{C}} := C^{a}_{\ a} = 0$ so finally the characteristic polynomial $\mathcal{W} (x)$ takes the form
\begin{equation}
\label{polynomial_quartic}
 \mathcal{W} (x) =  x^4 + \underset{[2]}{\mathbb{C}} x^2 - \underset{[3]}{\mathbb{C}} x + \underset{[4]}{\mathbb{C}}
\end{equation}
In $\mathbf{UR}$ coefficients $\underset{[i]}{\mathbb{C}} \in \mathbf{R}$. Criteria which allow us to distinguish the properties of the roots of $\mathcal{W} (x)$ have been widely discussed in \cite{Plebanski_Spinors, Rees, Lazard, Zhao}. Define 
\begin{eqnarray}
\label{kryteria}
&& -8J := \frac{1}{2} \underset{[3]}{\mathbb{C}}^{2} - \frac{4}{3} \underset{[2]}{\mathbb{C}} \underset{[4]}{\mathbb{C}} + \frac{1}{27} \underset{[2]}{\mathbb{C}}^{3} \ , \ \ \ I := \underset{[4]}{\mathbb{C}} + \frac{1}{12} \underset{[2]}{\mathbb{C}}^{2} \ , \ \ \ K:= \frac{1}{4} \underset{[3]}{\mathbb{C}} 
\\ \nonumber
&& L:= \frac{1}{6} \underset{[2]}{\mathbb{C}} \ , \ \ \ N := \frac{1}{4} \underset{[2]}{\mathbb{C}}^{2} - \underset{[4]}{\mathbb{C}} \ , \ \ \ P:=-9\underset{[3]}{\mathbb{C}}^{2} - 2 \underset{[2]}{\mathbb{C}} (\underset{[2]}{\mathbb{C}}^{2} - 4 \underset{[4]}{\mathbb{C}})
\end{eqnarray}
Then the discriminant of the polynomial (\ref{polynomial_quartic}) reads
\begin{equation}
\Delta = 256(I^{3} - 27J^{2})
\end{equation}
As it was mentioned in the previous subsection, there are exactly 9 cases which should be distinguished using the criteria from the Table \ref{kryteria_pierwiastkow_4stopnia}.
\begin{table}[!h]
\renewcommand{\arraystretch}{1.3}
\begin{center}
\begin{tabular}{|c|c|c|c|}   \hline
\multicolumn{3}{|c|}{Criteria}  & Roots  \\  \hline
\multicolumn{3}{|c|}{ $\Delta < 0$ }  & $R_{1}R_{2}Z \bar{Z}$ \\ \hline
$\Delta > 0$ & \multicolumn{2}{|c|}{ $L < 0$ \textrm{and} $N > 0$ } & $R_{1}R_{2}R_{3}R_{4}$ \\ \cline{2-4}
  & \multicolumn{2}{|c|}{ $L \geq 0$ \textrm{or} $N < 0$ } & $Z_{1}\bar{Z}_{1}Z_{2}\bar{Z}_{2}$ \\ \hline
  $\Delta = 0$ &  $I \ne 0$,  $J \ne 0$ & $P > 0$  & $R_{1}R_{2}R_{3}^{2}$ \\ \cline{3-4}
  & ($K \ne 0$ \textrm{or} $N\ne 0$) & $P < 0$ & $R^{2}Z\bar{Z} $ \\ \cline{2-4}
   &  $I \ne 0$,  $J \ne 0$ & $J < 0$  & $R_{1}^{2}R_{2}^2$ \\ \cline{3-4}
  & $K =N= 0$ & $J > 0$ & $Z^{2} \bar{Z}^{2} $ \\ \cline{2-4}
  &  $I =J= 0$ & $N\ne 0$ and $K \ne 0$  & $R_{1}R_{2}^{3}$ \\ \cline{3-4}
  &  & $N=K=0$ & $R^{4}  $ \\ \hline
\end{tabular}
\caption{Roots of the quartic equation with real coefficients.}
\label{kryteria_pierwiastkow_4stopnia}
\end{center}
\end{table}

In $\mathbf{CR}$ the coefficients $\underset{[i]}{\mathbb{C}}$ are complex. There are only 5 distinct cases (see Table \ref{kryteria_pierwiastkow_4stopnia_zespolony}).
\begin{table}[!h]
\renewcommand{\arraystretch}{1.3}
\begin{center}
\begin{tabular}{|c|c|c|c|}   \hline
\multicolumn{3}{|c|}{Criteria}  & Roots  \\  \hline
\multicolumn{3}{|c|}{ $\Delta \ne 0$ }  & $Z_{1}Z_{2}Z_{3}Z_{4}$ \\ \hline
  $\Delta = 0$ &  $I \ne 0$,  $J \ne 0$ & $P \ne 0$  & $Z_{1}Z_{2}Z_{3}^{2}$ \\ \cline{3-4}
  &  & $P =0$ & $Z_{1}^{2}Z_{2}^{2}$ \\ \cline{2-4}
  &  $I =J= 0$ & $L\ne 0$   & $Z_{1}Z_{2}^{3}$ \\ \cline{3-4}
  &  & $L=0$ & $Z^{4}  $ \\ \hline
\end{tabular}
\caption{Roots of the quartic equation with complex coefficients.}
\label{kryteria_pierwiastkow_4stopnia_zespolony}
\end{center}
\end{table}

\subsection{Terminology and symbols}

To classify traceless Ricci tensor in $\mathbf{UR}$ we use the notation similar to Plebański's notation from \cite{Plebanski_klasyfikacja_matter} and the Plebański - Przanowski notation from \cite{Przanowski_classification}. The number of eigenvectors are considered as a main criterion while the properties of the eigenvalues and the form of the minimal polynomial serve as sub criteria.

The complete information about the type of the matrix $(C^{a}_{\ b})$ is gathered in the symbol
\begin{equation}
\label{symbol_uzywany_przeznas}
^{[\textrm{A}]_{j} \otimes [\textrm{B}]_{j}}[n_{1} E_{1} - n_{2} E_{2} - ...]^{v}_{(q_{1}q_{2}...)}
\end{equation}
Inside the square bracket all different eigenvalues $E_{i}$, $i=1,2,...,N_{0}$ together with their multiplicities $n_{i}$ are listed. Of course
\begin{eqnarray}
&&n_{1} + n_{2} + ... + n_{N_{0}} =4
\\ \nonumber
&&n_{1} E_{1} + n_{2}E_{2} + ... + n_{N_{0}}E_{N_{0}} =0
\end{eqnarray}
The last equality follows from the fact, that the matrix $(C^{a}_{\ b})$ is traceless. The characteristic polynomial takes the form
\begin{equation}
\mathcal{W}(x) = \prod_{i=1}^{N_{0}} (x-E_{i})^{n_{i}}
\end{equation}

A complex eigenvalue is denoted by $Z$ and the real one by $R$. Real eigenvalues have additional superscript which denotes the type of the corresponding eigenvector. $R^{s}$ means, that the eigenvector which corresponds to the eigenvalue $R$ is spacelike, $R^{t}$ - timelike, $R^{n}$ - null, $R^{ns}$ - null or spacelike, $R^{nt}$ - null or timelike and finally $R^{nst}$ means, that the eigenvector can be of the arbitrary type. [With respect to the orthonormal tetrad (\ref{orthonormal_tetrad}) the definitions of spacelike, timelike and null vectors are as follows: $V^{a}V_{a} >0$ means, that $V^{a}$ is spacelike, $V^{a}V_{a} <0$ stands for a timelike vector and finally, $V^{a}V_{a} =0$ means, that the vector is null]. 

Superscript $v$ denotes the number of eigenvectors. Numbers $q_{i}$ in the round bracket determines the form of the minimal polynomial, i.e., the polynomial of the lowest possible order with the leading term equal $1$ such that $\mathcal{W}_{\textrm{min}} (C^{a}_{\ b})=0$. Namely, the minimal polynomial of the matrix $(C^{a}_{\ b})$ has the form
\begin{equation}
\mathcal{W}_{\textrm{min}} (x) =  \prod_{i=1}^{N_{0}} (x-E_{i})^{q_{i}}
\end{equation}

Finally, the symbol $[\textrm{A}]_{j} \otimes [\textrm{B}]_{j}$ defines the Petrov-Penrose types of the Plebański spinors, $V_{ABCD}$ and $V_{\dot{A}\dot{B}\dot{C}\dot{D}}$, respectively (\ref{Plebannnn_spinorrs}). For example, $[\textrm{III}]_{r} \otimes [\textrm{N}]_{r}$ means, that $V_{ABCD}$ is of the type $[\textrm{III}]_{r}$ while $V_{\dot{A}\dot{B}\dot{C}\dot{D}}$ is of the type $[\textrm{N}]_{r}$.


\setlength\arraycolsep{2pt}
\setcounter{equation}{0}

\renewcommand{\arraystretch}{1.5}

\section{Classification of the traceless Ricci tensor in $\mathbf{UR}$}
\label{Classification_of_the_traceless_Ricci_tensor}

\subsection{Parent Types}

The eigenvalue criteria (Table \ref{kryteria_pierwiastkow_4stopnia}), the number and the type of eigenvectors and the Petrov-Penrose type of the undotted and dotted Plebański spinors allow to distinguish exactly 33 types of the traceless Ricci tensor. They appear as the degenerations of 9 \textsl{parent Types} (according to Plebański's terminology, "Types" by capital "T"). Each of these parent Types has the minimal equation being exactly the Hamilton-Cayley equation. The symbols of the Types are quite similar like the symbols of Petrov-Penrose types of the Plebański spinors. To distinguish them we do not put the symbol of the Types into the square bracket (as we do in the case of the Petrov-Penrose types of the Plebański spinors). Types I and II have subscripts $r$ (all eigenvectors real), $c$ (all eigenvectors complex) or $rc$ (two eigenvectors complex, one or two eigenvectors real). Types III and IV have only real eigenvectors. However, Type III admits two null eigenvectors (subscript $n$), one null eigenvector and one timelike (subscript $t$) or one null eigenvector and one spacelike (subscript $s$). We use the symbols of the parent Types in the $\mathbf{CR}$ like in \cite{Przanowski_classification} (I,II,$\textrm{III}_{C}$,$\textrm{III}_{N}$ and IV).
\begin{table}[!h]
\begin{center}
\begin{tabular}{|c|c|}   \hline
Type  & Symbols of the parent Types  \\  \hline
 $\textrm{I}_{c}$ & $^{[\textrm{I}]_{c} \otimes [\textrm{I}]_{r}} [Z_{1}-\bar{Z}_{1}-Z_{2}-\bar{Z}_{2}]^{4}_{(1111)}$ , \ \ \ $^{[\textrm{I}]_{r} \otimes [\textrm{I}]_{c}} [Z_{1}-\bar{Z}_{1}-Z_{2}-\bar{Z}_{2}]^{4}_{(1111)}$
\\  \hline
 $\textrm{I}_{rc}$ & $^{[\textrm{I}]_{rc} \otimes [\textrm{I}]_{rc}} [Z-\bar{Z}-R_{1}^{s}-R_{2}^{t}]^{4}_{(1111)}$
\\  \hline
 $\textrm{I}_{r}$ & $^{[\textrm{I}]_{c} \otimes [\textrm{I}]_{c}} [R_{1}^{s}-R_{2}^{s}-R_{3}^{t}-R_{4}^{t}]^{4}_{(1111)} $   , \ \ \ $^{[\textrm{I}]_{r} \otimes [\textrm{I}]_{r}} [R_{1}^{s}-R_{2}^{s}-R_{3}^{t}-R_{4}^{t}]^{4}_{(1111)} $
\\  \hline
$ \textrm{II}_{rc}$ &  $^{[\textrm{II}]_{rc} \otimes [\textrm{II}]_{r}} [Z-\bar{Z}-2R^{n}]^{3}_{(112)}$  , \ \ \ $^{[\textrm{II}]_{r} \otimes [\textrm{II}]_{rc}} [Z-\bar{Z}-2R^{n}]^{3}_{(112)}$
\\  \hline
 $\textrm{II}_{r}$ &  $^{[\textrm{II}]_{rc} \otimes [\textrm{II}]_{rc}} [R_{1}^{s}-R_{2}^{t}-2R_{3}^{n}]^{3}_{(112)}$  , \ \ \ $^{[\textrm{II}]_{r} \otimes [\textrm{II}]_{r}} [R_{1}^{s}-R_{2}^{t}-2R_{3}^{n}]^{3}_{(112)}$
\\  \hline
 $\textrm{III}_{n}$ &  $^{[\textrm{D}]_{r} \otimes [\textrm{II}]_{r}} [2R_{1}^{n}-2R_{2}^{n}]^{2}_{(22)}$   , \ \ \  $^{[\textrm{II}]_{r} \otimes [\textrm{D}]_{r}} [2R_{1}^{n}-2R_{2}^{n}]^{2}_{(22)}$
\\  \hline
 $\textrm{III}_{s}$  &  $^{[\textrm{III}]_{r} \otimes [\textrm{III}]_{r}} [R_{1}^{s}-3R_{2}^{n}]^{2}_{(13)}$
\\  \hline
 $\textrm{III}_{t}$ &  $^{[\textrm{III}]_{r} \otimes [\textrm{III}]_{r}} [R_{1}^{t}-3R_{2}^{n}]^{2}_{(13)}$ 
\\  \hline
 $\textrm{IV}$ & $^{[\textrm{N}]_{r} \otimes [\textrm{III}]_{r}} [4R^{n}]^{1}_{(4)}$  , \ \ \  $^{[\textrm{III}]_{r} \otimes [\textrm{N}]_{r}} [4R^{n}]^{1}_{(4)}$
\\  \hline
\end{tabular}
\caption{Parent Types of $C_{ab}$.}
\label{Parent_Types}
\end{center}
\end{table}

In the next section we present the canonical forms of the parent Types and the tables of possible degenerations together with the continuous characteristics of the matrix $(C^{a}_{\ b})$. For the canonical forms we use both null and orthonormal tetrads. The classification of the traceless Ricci tensor in complex spaces can be treated as a "generic" classification for the $\mathbf{UR}$. This is why we list the Plebański - Przanowski's types described in details in \cite{Przanowski_classification} (we keep the original symbols of types used in \cite{Przanowski_classification}). For the Plebański - Przanowski classification we use the abbreviation \textsl{PP classification}.

\renewcommand{\arraystretch}{1.5}

\subsection{Type $\textrm{I}$ (4 eigenvectors)}

\subsubsection{Type $\textrm{I}_{r}$ (4 real eigenvectors; 2 spacelike and 2 timelike eigenvectors)}

The canonical form of $C_{ab}$ for the parent Type $\textrm{I}_{r}$ reads
\begin{eqnarray}
C_{ab} &=& R_{1}^{s} \, E_{1'a}E_{1'b} + R_{2}^{s} \, E_{2'a}E_{2'b} - R_{3}^{t} \, E_{3'a}E_{3'b} - R_{4}^{t} \, E_{4'a}E_{4'b}
\\ \nonumber
&=& \frac{1}{2} (R_{1}^{s}-R_{3}^{t}) (e_{1a}e_{1b}+e_{2a}e_{2b}) + \frac{1}{2} (R_{1}^{s}+R_{3}^{t}) (e_{1a}e_{2b}+e_{2a}e_{1b})
\\ \nonumber 
&&+ \frac{1}{2} (R_{2}^{s}-R_{4}^{t}) (e_{3a}e_{3b}+e_{4a}e_{4b}) + \frac{1}{2} (R_{2}^{s}+R_{4}^{t}) (e_{3a}e_{4b}+e_{4a}e_{3b})
\end{eqnarray}
The eigenvectors and corresponding eigenvalues are
\begin{displaymath}
E_{1'} \longleftrightarrow R_{1}^{s} \ , \ \ \ E_{2'} \longleftrightarrow R_{2}^{s} \ , \ \ \ 
E_{3'} \longleftrightarrow R_{3}^{t} \ , \ \ \ E_{4'} \longleftrightarrow R_{4}^{t} 
\end{displaymath}
The eigenvalues have to satisfy
\begin{displaymath}
R_{1}^{s}  + R_{2}^{s} + R_{3}^{t} + R_{4}^{t} =0 
\end{displaymath}
Using (\ref{relacje_pomocne_przy_spinorach_Plebanskiego}) one finds the form of the Plebański spinors
\begin{eqnarray}
V_{ABCD} &=& \frac{1}{2}(R_{1}^{s}-R_{3}^{t})(R_{2}^{s}-R_{4}^{t}) (k_{A}k_{B}k_{C}k_{D} + l_{A}l_{B}l_{C}l_{D}) 
\\ \nonumber
&& + \frac{1}{2} \big( (R_{2}^{s}-R_{4}^{t})^{2} - (3R_{1}^{s}+R_{3}^{t})(R_{1}^{s}+3R_{3}^{t}) \big) \, k_{(A}k_{B}l_{C}l_{D)}
\\ \nonumber
V_{\dot{A}\dot{B}\dot{C}\dot{D}} &=& \frac{1}{2}(R_{1}^{s}-R_{3}^{t})(R_{2}^{s}-R_{4}^{t}) (k_{\dot{A}}k_{\dot{B}}k_{\dot{C}}k_{\dot{D}} + l_{\dot{A}}l_{\dot{B}}l_{\dot{C}}l_{\dot{D}}) 
\\ \nonumber
&& + \frac{1}{2} \big( (R_{2}^{s}-R_{4}^{t})^{2} - (3R_{1}^{s}+R_{3}^{t})(R_{1}^{s}+3R_{3}^{t}) \big) \, k_{(\dot{A}}k_{\dot{B}}l_{\dot{C}}l_{\dot{D})}
\end{eqnarray}
Investigation of the polynomials $\mathcal{V} (z)$ and $\dot{\mathcal{V}} (\dot{z})$ (defined in the subsection \ref{subsekcja_Petrov-Penrose_classification}) proves, that both Plebański spinors are, in general, of the Petrov-Penrose types $[\textrm{I}]_{r}$ or $[\textrm{I}]_{c}$. Define the quantity $\sigma$ by the formula
\begin{equation}
\sigma_{1} := (R_{3}^{t}-R_{1}^{s})(R_{3}^{t}-R_{2}^{s})(R_{3}^{t}+R_{1}^{s}+2R_{2}^{s})(R_{3}^{t}+2R_{1}^{s}+R_{2}^{s})
\end{equation}
Then we get the criterion
\begin{eqnarray}
&& \sigma_{1} < 0 \ \Longleftrightarrow \ \textrm{ both } V_{ABCD} \textrm{ and } V_{\dot{A}\dot{B}\dot{C}\dot{D}} \textrm{ are of the type } [\textrm{I}]_{r} 
\\ \nonumber
&& \sigma_{1} > 0 \ \Longleftrightarrow \ \textrm{ both } V_{ABCD} \textrm{ and } V_{\dot{A}\dot{B}\dot{C}\dot{D}} \textrm{ are of the type } [\textrm{I}]_{c} 
\end{eqnarray}
\begin{table}[!h]
\begin{center}
\begin{tabular}{|c|c|c|c|}   \hline
\multicolumn{2}{|c|}{PP classification}  & \multicolumn{2}{|c|}{Neutral signature classification}  \\  \hline
Eigenvalues & Type $\textrm{I}$ & Eigenvalues & Type $\textrm{I}_{r}$ \\ \hline
  $Z_{1}Z_{2}Z_{3}Z_{4}$ & $[C_{1}-C_{2}-C_{3}-C_{4}]_{4}$ & $R_{1}R_{2}R_{3}R_{4}$ & $ ^{[\textrm{I}]_{r} \otimes [\textrm{I}]_{r}} [R_{1}^{s}-R_{2}^{s}-R_{3}^{t}-R_{4}^{t}]^{4}_{(1111)}$ \\ \cline{4-4}
  &  &  & $ ^{[\textrm{I}]_{c} \otimes [\textrm{I}]_{c}} [R_{1}^{s}-R_{2}^{s}-R_{3}^{t}-R_{4}^{t}]^{4}_{(1111)}$
  \\ \hline
 $Z_{1}Z_{2}Z_{3}^{2}$ & $[C_{1}-C_{2}-2N]_{3}$ & $R_{1}R_{2}R_{3}^{2}$ & $ ^{[\textrm{D}]_{c} \otimes [\textrm{D}]_{c}} [2R_{1}^{s}-R_{2}^{t}-R_{3}^{t}]^{4}_{(111)}$
  \\   \cline{4-4} 
   &  &  & $ ^{[\textrm{D}]_{r} \otimes [\textrm{D}]_{r}} [R_{1}^{s}-2R_{2}^{nst}-R_{3}^{t}]^{4}_{(111)}$
  \\  \cline{4-4} 
  &  &  & $ ^{[\textrm{D}]_{c} \otimes [\textrm{D}]_{c}} [R_{1}^{s}-R_{2}^{s}-2R_{3}^{t}]^{4}_{(111)}$
  \\ \hline
  $Z_{1}^{2}Z_{2}^{2}$ & $[2N_{1}-2N]_{2}$ & $R_{1}^{2}R_{2}^{2}$ & $ ^{[\textrm{D}]_{c} \otimes [\textrm{D}]_{c}} [2R_{1}^{s}-2R_{2}^{t}]^{4}_{(11)}$
  \\   \cline{4-4} 
   &  &  & $ ^{[\textrm{D}]_{r} \otimes [\textrm{D}]_{r}} [2R_{1}^{nst}-2R_{2}^{nst}]^{4}_{(11)}$
   \\ \hline
  $Z_{1}Z_{2}^{3}$ & $[C_{1}-3N]_{2}$ & $R_{1}R_{2}^{3}$ & $ ^{[-] \otimes [-]} [R_{1}^{s}-3R_{2}^{nst}]^{4}_{(11)}$
  \\   \cline{4-4} 
   &  &  & $ ^{[-] \otimes [-]} [R_{1}^{t}-3R_{2}^{nst}]^{4}_{(11)}$
   \\ \hline
   $Z^{4}$ & $[4N]_{1}$ & $R^{4}$ & $ ^{[-] \otimes [-]} [4R^{nst}]^{4}_{(1)}$
   \\ \hline
\end{tabular}
\caption{Subtypes of the Type $\textrm{I}_{r}$.}
\end{center}
\end{table}
\begin{Scheme}[!h]
\begin{displaymath}
\resizebox{1\textwidth}{!}{
\xymatrixcolsep{0in}
\xymatrixrowsep{1.3cm}
\xymatrix{
^{[\textrm{I}]_{c} \otimes [\textrm{I}]_{c}} [R_{1}^{s}-R_{2}^{s}-R_{3}^{t}-R_{4}^{t}]^{4}_{(1111)} \ar[d] \ar[dr] \ar[drr]  &   & ^{[\textrm{I}]_{r} \otimes [\textrm{I}]_{r}} [R_{1}^{s}-R_{2}^{s}-R_{3}^{t}-R_{4}^{t}]^{4}_{(1111)} \ar[d] & \\
^{[\textrm{D}]_{c} \otimes [\textrm{D}]_{c}} [2R_{1}^{s}-R_{2}^{t}-R_{3}^{t}]^{4}_{(111)} \ar[d] \ar[dr] & 
^{[\textrm{D}]_{c} \otimes [\textrm{D}]_{c}} [R_{1}^{s}-R_{2}^{s}-2R_{3}^{t}]^{4}_{(111)} \ar[dl] \ar[dr] & 
^{[\textrm{D}]_{r} \otimes [\textrm{D}]_{r}} [R_{1}^{s}-2R_{2}^{nst}-R_{3}^{t}]^{4}_{(111)} \ar[dl] \ar[d] \ar[dr] & \\
^{[\textrm{D}]_{c} \otimes [\textrm{D}]_{c}} [2R_{1}^{s}-2R_{2}^{t}]^{4}_{(11)} \ar[dr] & 
^{[-] \otimes [-]} [R_{1}^{t}-3R_{2}^{nst}]^{4}_{(11)} \ar[d] & 
^{[-] \otimes [-]} [R_{1}^{s}-3R_{2}^{nst}]^{4}_{(11)} \ar[dl] & 
^{[\textrm{D}]_{r} \otimes [\textrm{D}]_{r}} [2R_{1}^{nst}-2R_{2}^{nst}]^{4}_{(11)} \ar[dll] \\
& ^{[-] \otimes [-]} [4R^{nst}]^{4}_{(1)} & & &
}
}
\end{displaymath} 
\caption{Degeneration scheme of the Type $\textrm{I}_{r}$.}
\label{Degeneration_scheme_of_the_Type_I}
\end{Scheme}

\subsubsection{Type $\textrm{I}_{c}$ (4 complex eigenvectors)}

The canonical form of the $C_{ab}$ for the parent Type $\textrm{I}_{c}$ has the form
\begin{eqnarray}
C_{ab}  &=& \frac{1}{2} (Z_{1} +\bar{Z}_{1}) (E_{2'a}E_{2'b}-E_{4'a}E_{4'b}) + \frac{i}{2} (Z_{1} -\bar{Z}_{1}) (E_{2'a}E_{4'b}+E_{4'a}E_{2'b})
\\ \nonumber 
&& +\frac{1}{2} (Z_{2} +\bar{Z}_{2}) (E_{1'a}E_{1'b}-E_{3'a}E_{3'b}) + \frac{i}{2} (Z_{2} -\bar{Z}_{2}) (E_{1'a}E_{3'b}+E_{3'a}E_{1'b})
\\ \nonumber
&=& \frac{1}{2} (Z_{1} +\bar{Z}_{1}) (e_{3a}e_{4b}+e_{4a}e_{3b}) + \frac{i}{2} (Z_{1} -\bar{Z}_{1}) (e_{3a}e_{3b}-e_{4a}e_{4b})
\\ \nonumber 
&&+ \frac{1}{2} (Z_{2} +\bar{Z}_{2}) (e_{1a}e_{2b}+e_{2a}e_{1b}) + \frac{i}{2} (Z_{2} -\bar{Z}_{2}) (e_{1a}e_{1b}-e_{2a}e_{2b})
\end{eqnarray}
The eigenvectors and corresponding eigenvalues are
\begin{eqnarray}
\nonumber
&& \frac{1}{\sqrt{2}} (E_{2'}+iE_{4'}) \longleftrightarrow Z_{1} \ , \ \ \ 
\frac{1}{\sqrt{2}} (E_{2'}-iE_{4'}) \longleftrightarrow \bar{Z}_{1} 
\\ \nonumber
&& \frac{1}{\sqrt{2}} (E_{1'}+iE_{3'}) \longleftrightarrow Z_{2} \ , \ \ \ 
\frac{1}{\sqrt{2}} (E_{1'}-iE_{3'}) \longleftrightarrow \bar{Z}_{2} 
\end{eqnarray}
The constraints for the eigenvalues read
\begin{displaymath}
\textrm{Im}(Z_{1}) \ne 0 \ , \ \ \ \textrm{Im}(Z_{2}) \ne 0 \ , \ \ \ \textrm{Re}(Z_{1}) +\textrm{Re}(Z_{2}) = 0 
\end{displaymath}
The Plebański spinors are
\begin{eqnarray}
V_{ABCD} &=& 2  \textrm{Im}(Z_{1}) \textrm{Im} (Z_{2})  (k_{A}k_{B}k_{C}k_{D} + l_{A}l_{B}l_{C}l_{D}) 
\\ \nonumber
&& - \big( 2  (\textrm{Im}(Z_{1}))^{2} + 2  (\textrm{Im}(Z_{2}))^{2} + 8 (\textrm{Re} (Z_{1}))^{2} \big) \, k_{(A}k_{B}l_{C}l_{D)}
\\ \nonumber
V_{\dot{A}\dot{B}\dot{C}\dot{D}} &=& -2  \textrm{Im}(Z_{1}) \textrm{Im} (Z_{2})  (k_{\dot{A}}k_{\dot{B}}k_{\dot{C}}k_{\dot{D}} + l_{\dot{A}}l_{\dot{B}}l_{\dot{C}}l_{\dot{D}}) 
\\ \nonumber
&& - \big( 2  (\textrm{Im}(Z_{1}))^{2} + 2  (\textrm{Im}(Z_{2}))^{2} + 8 (\textrm{Re} (Z_{1}))^{2} \big) \, k_{(\dot{A}}k_{\dot{B}}l_{\dot{C}}l_{\dot{D})}
\end{eqnarray}
and they both are, in general, of the type $[\textrm{I}]_{r}$ and $[\textrm{I}]_{c}$. To distinguish these two types we have the following criterion
\begin{eqnarray}
&& \textrm{Im} (Z_{1}) \textrm{Im} (Z_{2})  < 0 \ \Longleftrightarrow \ V_{ABCD} \textrm{ is of the type } [\textrm{I}]_{c} \textrm{ and } V_{\dot{A}\dot{B}\dot{C}\dot{D}} \textrm{ is of the type } [\textrm{I}]_{r} \ \ \ \ \ \ \ \ \ 
\\ \nonumber
&& \textrm{Im} (Z_{1}) \textrm{Im} (Z_{2})  > 0 \ \Longleftrightarrow \ V_{ABCD} \textrm{ is of the type } [\textrm{I}]_{r} \textrm{ and } V_{\dot{A}\dot{B}\dot{C}\dot{D}} \textrm{ is of the type } [\textrm{I}]_{c}
\end{eqnarray}
It is interesting to note, that only the subtype $^{[\textrm{I}]_{r} \otimes [\textrm{I}]_{c}} [Z_{1}-\bar{Z}_{1}-Z_{2}-\bar{Z}_{2}]^{4}_{(1111)}$ allows the degeneration into the type $^{[\textrm{D}]_{r} \otimes [\textrm{D}]_{c}} [2Z-2\bar{Z}]^{4}_{(11)}$.
\begin{table}[!h]
\begin{center}
\begin{tabular}{|c|c|c|c|}   \hline
\multicolumn{2}{|c|}{PP classification}  & \multicolumn{2}{|c|}{Neutral signature classification}  \\  \hline
Eigenvalues & Type $\textrm{I}$ & Eigenvalues & Type $\textrm{I}_{c}$ \\ \hline
  $Z_{1}Z_{2}Z_{3}Z_{4}$ & $[C_{1}-C_{2}-C_{3}-C_{4}]_{4}$ & $Z_{1}\bar{Z}_{1}Z_{2}\bar{Z}_{2}$ & $ ^{[\textrm{I}]_{c} \otimes [\textrm{I}]_{r}} [Z_{1}-\bar{Z}_{1}-Z_{2}-\bar{Z}_{2}]^{4}_{(1111)}$
  \\ \cline{4-4}
    & &  & $ ^{[\textrm{I}]_{r} \otimes [\textrm{I}]_{c}} [Z_{1}-\bar{Z}_{1}-Z_{2}-\bar{Z}_{2}]^{4}_{(1111)}$
  \\ \hline
  $Z_{1}^{2}Z_{2}^{2}$ & $[2N_{1}-2N]_{2}$ & $Z^{2} \bar{Z}^{2} $ & $ ^{[\textrm{D}]_{r} \otimes [\textrm{D}]_{c}} [2Z-2\bar{Z}]^{4}_{(11)}$
   \\ \hline
\end{tabular}
\caption{Subtypes of the Type $\textrm{I}_{c}$.}
\end{center}
\end{table}
\begin{Scheme}[!h]
\begin{displaymath}
\xymatrixcolsep{1.0cm}
\xymatrixrowsep{1.0cm}
\xymatrix{
^{[\textrm{I}]_{c} \otimes [\textrm{I}]_{r}} [Z_{1}-\bar{Z}_{1}-Z_{2}-\bar{Z}_{2}]^{4}_{(1111)}  &
^{[\textrm{I}]_{r} \otimes [\textrm{I}]_{c}} [Z_{1}-\bar{Z}_{1}-Z_{2}-\bar{Z}_{2}]^{4}_{(1111)} \ar[d] \\
 & ^{[\textrm{D}]_{r} \otimes [\textrm{D}]_{c}} [2Z-2\bar{Z}]^{4}_{(11)}
}
\end{displaymath}
\caption{Degeneration scheme of the Type $\textrm{I}_{c}$.}
\end{Scheme}

\subsubsection{Type $\textrm{I}_{rc}$ (4 eigenvectors; two complex, one timelike and one spacelike eigenvectors)}

The canonical form of the $C_{ab}$ for the parent Type $\textrm{I}_{rc}$ is
\begin{eqnarray}
C_{ab}  &=& R_{1}^{s} E_{1'a}E_{1'b} - R_{2}^{t} E_{3'a}E_{3'b}
\\ \nonumber 
&& +\frac{1}{2} (Z +\bar{Z}) (E_{2'a}E_{2'b}-E_{4'a}E_{4'b}) + \frac{i}{2} (Z -\bar{Z}) (E_{2'a}E_{4'b}+E_{4'a}E_{2'b})
\\ \nonumber
&=& \frac{1}{2} (R_{1}^{s} -R_{2}^{t}) (e_{1a}e_{1b}+e_{2a}e_{2b}) + \frac{1}{2} (R_{1}^{s} +R_{2}^{t}) (e_{1a}e_{2b}+e_{2a}e_{1b})
\\ \nonumber 
&&+ \frac{1}{2} (Z +\bar{Z}) (e_{3a}e_{4b}+e_{4a}e_{3b}) + \frac{i}{2} (Z -\bar{Z}) (e_{3a}e_{3b}-e_{4a}e_{4b})
\end{eqnarray}
The eigenvectors and corresponding eigenvalues are
\begin{displaymath}
E_{1'} \longleftrightarrow R_{1}^{s} \ , \ \ \ 
E_{3'} \longleftrightarrow R_{2}^{t} \ , \ \ \ 
\frac{1}{\sqrt{2}} (E_{2'}+iE_{4'}) \longleftrightarrow Z \ , \ \ \ 
\frac{1}{\sqrt{2}} (E_{2'}-iE_{4'}) \longleftrightarrow \bar{Z} 
\end{displaymath}
The relations between eigenvalues read
\begin{displaymath}
\textrm{Im}(Z) \ne 0 \ , \ \ \ R_{1}^{s}+R_{2}^{t} +2\textrm{Re}(Z) = 0 
\end{displaymath}
Plebański spinors have the following form
\begin{eqnarray}
V_{ABCD} &=&   (R_{1}^{s}-R_{2}^{t}) \textrm{Im} (Z)  (k_{A}k_{B}k_{C}k_{D} - l_{A}l_{B}l_{C}l_{D}) 
\\ \nonumber
&& + \frac{1}{2} \big( (R_{1}^{s}-R_{2}^{t})^{2} - 4  (\textrm{Im}(Z))^{2} - 16 (\textrm{Re} (Z))^{2} \big) \, k_{(A}k_{B}l_{C}l_{D)}
\\ \nonumber
V_{\dot{A}\dot{B}\dot{C}\dot{D}} &=& (R_{1}^{s}-R_{2}^{t}) \textrm{Im} (Z)  (k_{\dot{A}}k_{\dot{B}}k_{\dot{C}}k_{\dot{D}} - l_{\dot{A}}l_{\dot{B}}l_{\dot{C}}l_{\dot{D}}) 
\\ \nonumber
&& + \frac{1}{2} \big( (R_{1}^{s}-R_{2}^{t})^{2} - 4 (\textrm{Im}(Z))^{2} - 16 (\textrm{Re} (Z))^{2} \big) \, k_{(\dot{A}}k_{\dot{B}}l_{\dot{C}}l_{\dot{D})}
\end{eqnarray}
Both Plebański spinors for nondegenerate Type $\textrm{I}_{rc}$ are of the Petrov-Penrose type $[\textrm{I}]_{rc}$.
\begin{table}[!h]
\begin{center}
\begin{tabular}{|c|c|c|c|}   \hline
\multicolumn{2}{|c|}{PP classification}  & \multicolumn{2}{|c|}{Neutral signature classification}  \\  \hline
Eigenvalues & Type $\textrm{I}$ & Eigenvalues & Type $\textrm{I}_{RC}$ \\ \hline
  $Z_{1}Z_{2}Z_{3}Z_{4}$ & $[C_{1}-C_{2}-C_{3}-C_{4}]_{4}$ & $Z \bar{Z} R_{1}R_{2}$ & $ ^{[\textrm{I}]_{rc} \otimes [\textrm{I}]_{rc}} [Z-\bar{Z}-R_{1}^{s}-R_{2}^{t}]^{4}_{(1111)}$
  \\ \hline
  $Z_{1}Z_{2}Z_{3}^{2}$ & $[C_{1}-C_{2}-2N]_{2}$ & $Z \bar{Z}R^{2} $ & $ ^{[\textrm{D}]_{r} \otimes [\textrm{D}]_{r}} [Z-\bar{Z}-2R^{nst}]^{4}_{(111)}$
   \\ \hline
\end{tabular}
\caption{Subtypes of the Type $\textrm{I}_{rc}$.}
\end{center}
\end{table}
\begin{Scheme}[!h]
\begin{displaymath}
\xymatrixcolsep{0in}
\xymatrixrowsep{1.0cm}
\xymatrix{
^{[\textrm{I}]_{rc} \otimes [\textrm{I}]_{rc}} [Z-\bar{Z}-R_{1}^{s}-R_{2}^{t}]^{4}_{(1111)} \ar[d] \\
^{[\textrm{D}]_{r} \otimes [\textrm{D}]_{r}} [Z-\bar{Z}-2R^{nst}]^{4}_{(111)}
}
\end{displaymath}
\caption{Degeneration scheme of the Type $\textrm{I}_{rc}$.}
\end{Scheme}

\subsection{Type $\textrm{II}$ (3 eigenvectors)}

\subsubsection{Type $\textrm{II}_{r}$ (3 eigenvectors; one timelike, one spacelike and one null eigenvectors)}

The canonical form of the $C_{ab}$ for the parent Type $\textrm{II}_{r}$ is
\begin{eqnarray}
C_{ab}  &=& R_{1}^{s} E_{1'a}E_{1'b} - R_{2}^{t} E_{3'a}E_{3'b}
\\ \nonumber 
&& +R_{3}^{n} (E_{2'a}E_{2'b}-E_{4'a}E_{4'b}) + \frac{1}{2} (E_{2'a}E_{2'b}+E_{4'a}E_{4'b} - E_{2'a}E_{4'b} - E_{4'a}E_{2'b} )
\\ \nonumber
&=& \frac{1}{2} (R_{1}^{s} -R_{2}^{t}) (e_{1a}e_{1b}+e_{2a}e_{2b}) + \frac{1}{2} (R_{1}^{s} +R_{2}^{t}) (e_{1a}e_{2b}+e_{2a}e_{1b})
\\ \nonumber 
&&+ R_{3}^{n} (e_{3a}e_{4b}+e_{4a}e_{3b}) +  e_{4a}e_{4b}
\end{eqnarray}
The eigenvectors and corresponding eigenvalues are
\begin{displaymath}
E_{1'} \longleftrightarrow R_{1}^{s} \ , \ \ \ 
E_{3'} \longleftrightarrow R_{2}^{t} \ , \ \ \ 
\frac{1}{\sqrt{2}} (E_{2'}- E_{4'}) \longleftrightarrow R_{3}^{n} 
\end{displaymath}
The eigenvalues have to satisfy the relation
\begin{displaymath}
R_{1}^{s}+R_{2}^{t}+2R_{3}^{n}=0
\end{displaymath}
Plebański spinors for the Type $\textrm{II}_{r}$ can be brought to the form
\begin{eqnarray}
V_{ABCD} &=& \frac{1}{2} \Big( 2(R_{1}^{s}-R_{2}^{t}) \, k_{(A}k_{B} -  (3R_{1}^{s}+R_{2}^{t})(R_{1}^{s}+3R_{2}^{t}) \, l_{(A}l_{B} \Big) \, k_{C}k_{D)}
\\ \nonumber
V_{\dot{A}\dot{B}\dot{C}\dot{D}} &=& \frac{1}{2} \Big( 2(R_{1}^{s}-R_{2}^{t}) \, k_{(\dot{A}}k_{\dot{B}} -  (3R_{1}^{s}+R_{2}^{t})(R_{1}^{s}+3R_{2}^{t}) \, l_{(\dot{A}}l_{\dot{B}} \Big) \, k_{\dot{C}}k_{\dot{D})}
\end{eqnarray}
Consider the quantity $\sigma_{2}$
\begin{equation}
\sigma_{2} := (R_{1}^{s}-R_{2}^{t})(3R_{1}^{s}+R_{2}^{t})(R_{1}^{s}+3R_{2}^{t})
\end{equation}
Then we find the following criterion
\begin{eqnarray}
&& \sigma_{2} > 0 \ \Longleftrightarrow \ V_{ABCD} \textrm{ and } V_{\dot{A}\dot{B}\dot{C}\dot{D}} \textrm{ are of the type } [\textrm{II}]_{r}
\\ \nonumber
&& \sigma_{2} < 0 \ \Longleftrightarrow \ V_{ABCD} \textrm{ and } V_{\dot{A}\dot{B}\dot{C}\dot{D}} \textrm{ are of the type } [\textrm{II}]_{rc}
\end{eqnarray}
\begin{table}[!h]
\begin{center}
\begin{tabular}{|c|c|c|c|}   \hline
\multicolumn{2}{|c|}{PP classification}  & \multicolumn{2}{|c|}{Neutral signature classification}  \\  \hline
Eigenvalues & Type $\textrm{II}$ & Eigenvalues & Type $\textrm{II}_{r}$ \\ \hline
  $Z_{1}Z_{2}Z_{3}^{2}$ & $[C_{1}-C_{2}-2N]_{4}$ & $R_{1}R_{2}R_{3}^{2}$ & $ ^{[\textrm{II}]_{rc} \otimes [\textrm{II}]_{rc}} [R_{1}^{s}-R_{2}^{t}-2R_{3}^{n}]^{3}_{(112)}$
  \\ \cline{4-4}
   &  &  & $ ^{[\textrm{II}]_{r} \otimes [\textrm{II}]_{r}} [R_{1}^{s}-R_{2}^{t}-2R_{3}^{n}]^{3}_{(112)}$
  \\ \hline
  $Z_{1}^{2}Z_{2}^{2}$ & $[2N_{1}-2N]_{(1-2)}$ & $R_{1}^{2}R_{2}^2$ & $ ^{[\textrm{D}]_{r} \otimes [\textrm{D}]_{r}} [2R_{1}^{nst}-2R_{2}^{n}]^{3}_{(12)}$
  \\ \hline 
  $Z_{1}Z_{2}^{3}$ & $[C_{1}-3N]_{3}$ & $R_{1}R_{2}^{3}$ & $ ^{[\textrm{N}]_{r} \otimes [\textrm{N}]_{r}} [R_{1}^{s}-3R_{2}^{nt}]^{3}_{(12)}$
  \\   \cline{4-4} 
   &  &  & $ ^{[\textrm{N}]_{r} \otimes [\textrm{N}]_{r}} [R_{1}^{t}-3R_{2}^{ns}]^{3}_{(12)}$ \\ \hline
   $Z^{4}$ & $^{(3)} [4N]_{2}$ & $R^{4}$ & $ ^{[-] \otimes [-]} [4R^{nst}]^{3}_{(2)}$
   \\ \hline
\end{tabular}
\caption{Subtypes of the Type $\textrm{II}_{r}$.}
\end{center}
\end{table}
\begin{Scheme}[!h]
\begin{displaymath}
\xymatrixcolsep{0in}
\xymatrixrowsep{1.3cm}
\xymatrix{
^{[\textrm{II}]_{rc} \otimes [\textrm{II}]_{rc}} [R_{1}^{s}-R_{2}^{t}-2R_{3}^{n}]^{3}_{(112)} \ar[d] \ar[dr] \ar[drr] &   & ^{[\textrm{II}]_{r} \otimes [\textrm{II}]_{r}} [R_{1}^{s}-R_{2}^{t}-2R_{3}^{n}]^{3}_{(112)} \ar[dll] \ar[dl] \ar[d] \\
^{[\textrm{D}]_{r} \otimes [\textrm{D}]_{r}} [2R_{1}^{nst}-2R_{2}^{n}]^{3}_{(12)} \ar[dr] &
^{[\textrm{N}]_{r} \otimes [\textrm{N}]_{r}} [R_{1}^{s}-3R_{2}^{nt}]^{3}_{(12)}  \ar[d]  &
^{[\textrm{N}]_{r} \otimes [\textrm{N}]_{r}} [R_{1}^{t}-3R_{2}^{ns}]^{3}_{(12)}  \ar[dl] \\
&  ^{[-] \otimes [-]} [4R^{nst}]^{3}_{(2)} &
}
\end{displaymath}
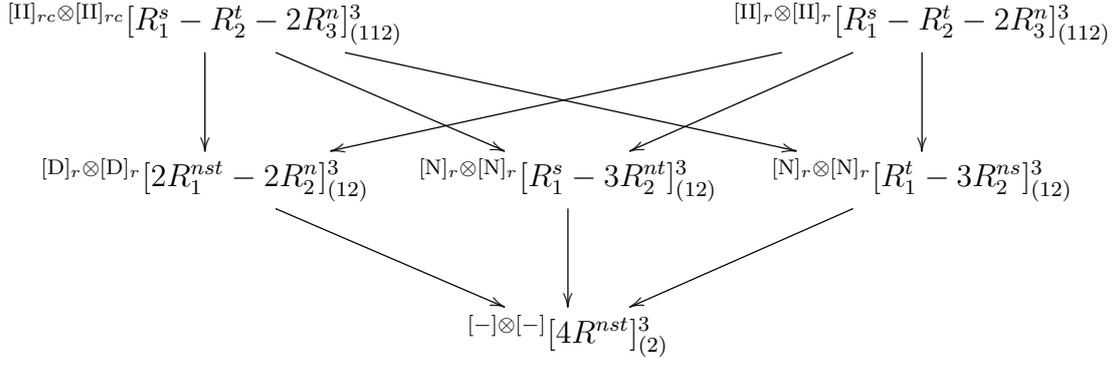
\caption{Degeneration scheme of the Type $\textrm{II}_{r}$.}
\end{Scheme}

\subsubsection{Type $\textrm{II}_{rc}$ (3 eigenvectors; two complex and one null eigenvectors)}

The canonical form of the $C_{ab}$ for the parent Type $\textrm{II}_{rc}$ has the form
\begin{eqnarray}
C_{ab}  &=& \frac{1}{2} (Z +\bar{Z}) (E_{1'a}E_{1'b}-E_{3'a}E_{3'b}) + \frac{i}{2} (Z -\bar{Z}) (E_{1'a}E_{3'b}+E_{3'a}E_{1'b})
\\ \nonumber 
&& +R^{n} (E_{2'a}E_{2'b}-E_{4'a}E_{4'b}) + \frac{1}{2} (E_{2'a}E_{2'b}+E_{4'a}E_{4'b} - E_{2'a}E_{4'b} - E_{4'a}E_{2'b})
\\ \nonumber
&=& \frac{1}{2} (Z +\bar{Z}) (e_{1a}e_{2b}+e_{2a}e_{1b}) + \frac{i}{2} (Z -\bar{Z}) (e_{1a}e_{1b}-e_{2a}e_{2b})
\\ \nonumber 
&&+ R^{n} (e_{3a}e_{4b}+e_{4a}e_{3b}) +    e_{4a}e_{4b}
\end{eqnarray}
The eigenvectors and corresponding eigenvalues are given by
\begin{displaymath}
\frac{1}{\sqrt{2}} (E_{1'}+iE_{3'}) \longleftrightarrow Z \ , \ \ \ 
\frac{1}{\sqrt{2}} (E_{1'}-iE_{3'}) \longleftrightarrow \bar{Z} \ , \ \ \ 
\frac{1}{\sqrt{2}} (E_{2'} - E_{4'}) \longleftrightarrow R^{n}
\end{displaymath}
The conditions for eigenvalues are
\begin{displaymath}
\textrm{Im}(Z) \ne 0 \ , \ \ \ R^{n} +\textrm{Re}(Z) = 0 
\end{displaymath}
Plebański spinors read
\begin{eqnarray}
V_{ABCD} &=& 2 \Big( \textrm{Im} (Z) \, k_{(A}k_{B} -  \big( (\textrm{Im} (Z))^{2} +4 (\textrm{Re} (Z))^{2}  \big) \, l_{(A}l_{B} \Big) \, k_{C}k_{D)}
\\ \nonumber
V_{\dot{A}\dot{B}\dot{C}\dot{D}} &=& 2 \Big( - \textrm{Im} (Z) \, k_{(\dot{A}}k_{\dot{B}} -  \big( (\textrm{Im} (Z))^{2} +4 (\textrm{Re} (Z))^{2}  \big) \, l_{(\dot{A}}l_{\dot{B}} \Big) \, k_{\dot{C}}k_{\dot{D})}
\end{eqnarray}
This time we find the criterion
\begin{eqnarray}
&& \textrm{Im} (Z) > 0 \ \Longleftrightarrow \ V_{ABCD} \textrm{ is of the type } [\textrm{II}]_{r} \textrm{ and } V_{\dot{A}\dot{B}\dot{C}\dot{D}} \textrm{ is of the type } [\textrm{II}]_{rc} \ \ \ \ \ \ \ \ \ 
\\ \nonumber
&& \textrm{Im} (Z) < 0 \ \Longleftrightarrow \ V_{ABCD} \textrm{ is of the type } [\textrm{II}]_{rc} \textrm{ and } V_{\dot{A}\dot{B}\dot{C}\dot{D}} \textrm{ is of the type } [\textrm{II}]_{r}
\end{eqnarray}
\begin{table}[!h]
\begin{center}
\begin{tabular}{|c|c|c|c|}   \hline
\multicolumn{2}{|c|}{PP classification}  & \multicolumn{2}{|c|}{Neutral signature classification}  \\  \hline
Eigenvalues & Type $\textrm{II}$ & Eigenvalues & Type $\textrm{II}_{rc}$ \\ \hline
  $Z_{1}Z_{2}Z_{3}^{2}$ & $[C_{1}-C_{2}-2N]_{4}$ & $Z\bar{Z} R^{2}$ & $ ^{[\textrm{II}]_{r} \otimes [\textrm{II}]_{rc}} [Z-\bar{Z}-2R^{n}]^{3}_{(112)}$
  \\ \cline{4-4}
  &  &  & $ ^{[\textrm{II}]_{rc} \otimes [\textrm{II}]_{r}} [Z-\bar{Z}-2R^{n}]^{3}_{(112)}$
  \\ \hline
\end{tabular}
\caption{Subtypes of the Type $\textrm{II}_{rc}$.}
\end{center}
\end{table}

\subsection{Type $\textrm{III}$ (2 eigenvectors)}

\subsubsection{Types $\textrm{III}_{s}$ and $\textrm{III}_{t}$ (2 eigenvectors; one null and one spacelike or timelike eigenvectors)}

The canonical form of the $C_{ab}$ for the parent Type $\textrm{III}_{t}$ is
\begin{eqnarray}
C_{ab}  &=& -R_{1}^{t} E_{3'a}E_{3'b} + R_{2}^{n} (E_{1'a}E_{1'b}+E_{2'a}E_{2'b}-E_{4'a}E_{4'b}) 
\\ \nonumber
&&+ E_{1'a}E_{2'b}+E_{2'a}E_{1'b}
-E_{1'a}E_{4'b}-E_{4'a}E_{1'b}
\\ \nonumber 
&=& \frac{1}{2} (R_{2}^{n} + R_{1}^{t}) (e_{1a}e_{2b}+e_{2a}e_{1b}) + \frac{1}{2} (R_{2}^{n} - R_{1}^{t}) (e_{1a}e_{1b}+e_{2a}e_{2b})
\\ \nonumber 
&&+ R_{2}^{n} (e_{3a}e_{4b}+e_{4a}e_{3b}) + e_{1a}e_{4b}+ e_{4a}e_{1b}+ e_{2a}e_{4b}+ e_{4a}e_{2b}
\end{eqnarray}
The eigenvectors and corresponding eigenvalues are
\begin{displaymath}
E_{3'} \longleftrightarrow R_{1}^{t} \ , \ \ \ 
e_{4} \longleftrightarrow R_{2}^{n}
\end{displaymath}
The eigenvalues have to satisfy the condition
\begin{displaymath}
R_{1}^{t}+3R_{2}^{n}=0
\end{displaymath}
The Plebański spinors have the following form
\begin{eqnarray}
V_{ABCD} &=& -2 \big( k_{(A} + 8R_{2}^{n} \, l_{(A}    \big) \, k_{B}k_{C}k_{D)}
\\ \nonumber
V_{\dot{A}\dot{B}\dot{C}\dot{D}} &=& -2 \big( k_{(\dot{A}} + 8R_{2}^{n} \, l_{(\dot{A}}    \big) \, k_{\dot{B}}k_{\dot{C}}k_{\dot{D})}
\end{eqnarray}
For the nondegenerate Type $\textrm{III}_{t}$ these spinors are both of the Petrov-Penrose type $[\textrm{III}]_{r}$.

The canonical form of the $C_{ab}$ for the parent Type $\textrm{III}_{s}$ reads
\begin{eqnarray}
C_{ab}  &=& R_{1}^{s} E_{1'a}E_{1'b} + R_{2}^{n} (E_{2'a}E_{2'b}-E_{3'a}E_{3'b}-E_{4'a}E_{4'b}) 
\\ \nonumber
&&+ E_{3'a}E_{2'b}+E_{2'a}E_{3'b}
-E_{4'a}E_{3'b}-E_{3'a}E_{4'b}
\\ \nonumber 
&=& \frac{1}{2} (R_{1}^{s} - R_{2}^{n}) (e_{1a}e_{1b}+e_{2a}e_{2b}) + \frac{1}{2} (R_{1}^{s} + R_{2}^{n}) (e_{1a}e_{2b}+e_{2a}e_{1b})
\\ \nonumber 
&&+ R_{2}^{n} (e_{3a}e_{4b}+e_{4a}e_{3b}) + e_{1a}e_{4b}+ e_{4a}e_{1b}- e_{2a}e_{4b}- e_{4a}e_{2b}
\end{eqnarray}
The eigenvectors and corresponding eigenvalues are
\begin{displaymath}
E_{1'} \longleftrightarrow R_{1}^{s} \ , \ \ \ 
e_{4} \longleftrightarrow R_{2}^{n}
\end{displaymath}
The eigenvalues satisfy the relation
\begin{displaymath}
R_{1}^{s}+3R_{2}^{n}=0
\end{displaymath}
The Plebański spinors read
\begin{eqnarray}
V_{ABCD} &=& -2 \big( k_{(A} - 8R_{2}^{n} \, l_{(A}    \big) \, k_{B}k_{C}k_{D)}
\\ \nonumber
V_{\dot{A}\dot{B}\dot{C}\dot{D}} &=& -2 \big( k_{(\dot{A}} + 8R_{2}^{n} \, l_{(\dot{A}}    \big) \, k_{\dot{B}}k_{\dot{C}}k_{\dot{D})}
\end{eqnarray}
and they represent the Petrov-Penrose type $[\textrm{III}]_{r}$.
\begin{table}[!h]
\begin{center}
\begin{tabular}{|c|c|c|c|c|}   \hline
\multicolumn{2}{|c|}{PP classification}  & \multicolumn{3}{|c|}{Neutral signature classification}  \\  \hline
E. values & Type $\textrm{III}_{C}$ & E. values & Type $\textrm{III}_{t}$ & Type $\textrm{III}_{s}$\\ \hline
  $Z_{1}Z_{2}^{3}$ & $[C_{1}-3N]_{4}$ & $R_{1}R_{2}^{3}$ & $ ^{[\textrm{III}]_{r} \otimes [\textrm{III}]_{r}} [R_{1}^{t}-3R_{2}^{n}]^{2}_{(13)}$  & $ ^{[\textrm{III}]_{r} \otimes [\textrm{III}]_{r}} [R_{1}^{s}-3R_{2}^{n}]^{2}_{(13)}$
  \\ \hline
  $Z^{4}$ & $[4N]_{3}$ & $R^{4}$ & $ ^{[\textrm{N}]_{r} \otimes [\textrm{N}]_{r}} [4R^{nt}]^{2}_{(3)}$  & $ ^{[\textrm{N}]_{r} \otimes [\textrm{N}]_{r}} [4R^{ns}]^{2}_{(3)}$
   \\ \hline
\end{tabular}
\caption{Subtypes of the Types $\textrm{III}_{t}$ and $\textrm{III}_{s}$.}
\end{center}
\end{table}
\begin{Scheme}[!h]
\begin{displaymath}
\xymatrixcolsep{0in}
\xymatrixrowsep{1.0cm}
\xymatrix{
^{[\textrm{III}]_{r} \otimes [\textrm{III}]_{r}} [R_{1}^{t}-3R_{2}^{n}]^{2}_{(13)} \ar[d] \ \ \ \ \ \ \ \ \ &
^{[\textrm{III}]_{r} \otimes [\textrm{III}]_{r}} [R_{1}^{s}-3R_{2}^{n}]^{2}_{(13)} \ar[d] \\
^{[\textrm{N}]_{r} \otimes [\textrm{N}]_{r}} [4R^{nt}]^{2}_{(3)} \ \ \ \ \ \ \ \ \ &
^{[\textrm{N}]_{r} \otimes [\textrm{N}]_{r}} [4R^{ns}]^{2}_{(3)}
}
\end{displaymath}
\caption{Degeneration scheme of the Types $\textrm{III}_{t}$ and $\textrm{III}_{s}$.}
\end{Scheme}

\subsubsection{Type $\textrm{III}_{n}$ (2 eigenvectors; both null)}

We find here two subtypes. The canonical form of the $C_{ab}$ reads
\begin{equation}
C_{ab}  = e_{1a} e_{1b} + e_{4a} e_{4b} + R_{1}^{n} (e_{3a}e_{4b} + e_{4a} e_{3b} ) + R_{2}^{n} (e_{1a}e_{2b} + e_{2a} e_{1b} )
\end{equation}
The eigenvectors and corresponding eigenvalues are
\begin{displaymath}
e_{4} \longleftrightarrow R_{1}^{n} \ , \ \ \ 
e_{1} \longleftrightarrow R_{2}^{n}
\end{displaymath}
For the first subtype the Plebański spinors read
\begin{eqnarray}
V_{ABCD} &=& -8 (R_{1}^{n})^{2} \, k_{(A}k_{B}l_{C}l_{D)}
\\ \nonumber
V_{\dot{A}\dot{B}\dot{C}\dot{D}} &=& 2 \, (k_{(\dot{A}} + 2R_{1}^{n} \, l_{(\dot{A}}) (k_{\dot{B}} - 2R_{1}^{n} \, l_{\dot{B}})  k_{\dot{C}}k_{\dot{D})}
\end{eqnarray}
Undotted Plebański spinor for the first subtype of the Type $\textrm{III}_{n}$ is of the type $[\textrm{D}]_{r}$ and the dotted one is of the type $[\textrm{II}]_{r}$.

The second possibility is
\begin{equation}
C_{ab}  = e_{2a} e_{2b} + e_{4a} e_{4b} + R_{1}^{n} (e_{3a}e_{4b} + e_{4a} e_{3b} ) + R_{2}^{n} (e_{1a}e_{2b} + e_{2a} e_{1b} )
\end{equation}
The eigenvectors and corresponding eigenvalues are given by
\begin{displaymath}
e_{4} \longleftrightarrow R_{1}^{n} \ , \ \ \ 
e_{2} \longleftrightarrow R_{2}^{n}
\end{displaymath}
The second subtype is characterized by the following Plebański spinors
\begin{eqnarray}
V_{ABCD} &=& 2 \, (k_{(A} + 2R_{1}^{n} \, l_{(A}) (k_{B} - 2R_{1}^{n} \, l_{B})  k_{C}k_{D)}
\\ \nonumber
V_{\dot{A}\dot{B}\dot{C}\dot{D}} &=& -8 (R_{1}^{n})^{2} \, k_{(\dot{A}}k_{\dot{B}}l_{\dot{C}}l_{\dot{D})}
\end{eqnarray}
and they are of the Petrov-Penrose types $[\textrm{II}]_{r}$ and $[\textrm{D}]_{r}$, respectively.

The eigenvalues in both subtypes have to satisfy the relation
\begin{displaymath}
R_{1}^{n}+R_{2}^{n}=0
\end{displaymath}
\begin{table}[!h]
\begin{center}
\begin{tabular}{|c|c|c|c|}   \hline
\multicolumn{2}{|c|}{PP classification}  & \multicolumn{2}{|c|}{Neutral signature classification}  \\  \hline
Eigenvalues & Type $\textrm{III}_{N}$ & Eigenvalues & Type $\textrm{III}_{n}$ \\ \hline
  $Z_{1}^{2}Z_{2}^{2}$ & $[2N_{1}-2N]^{a}_{4}$ & $R_{1}^{2}R_{2}^{2}$ & $ ^{[\textrm{D}]_{r} \otimes [\textrm{II}]_{r}} [2R_{1}^{n}-2R_{2}^{n}]^{2}_{(22)}$
  \\ \cline{2-2}  \cline{4-4} 
  & $[2N_{1}-2N]^{b}_{4}$ &  & $ ^{[\textrm{II}]_{r} \otimes [\textrm{D}]_{r}} [2R_{1}^{n}-2R_{2}^{n}]^{2}_{(22)}$
  \\ \hline
  $Z^{4}$ & $^{(2)} [4N]^{a}_{2}$ & $R^{4}$ & $ ^{[-] \otimes [\textrm{N}]_{r}} [4R^{n}]^{2}_{(2)}$
  \\ \cline{2-2}  \cline{4-4} 
  & $^{(2)} [4N]^{b}_{2}$ &  & $ ^{[\textrm{N}]_{r} \otimes [-]} [4R^{n}]^{2}_{(2)}$ \\ \hline
\end{tabular}
\caption{Subtypes of the Type $\textrm{III}_{n}$.}
\end{center}
\end{table}
\begin{Scheme}[!h]
\begin{displaymath}
\xymatrixcolsep{0in}
\xymatrixrowsep{1.0cm}
\xymatrix{
^{[\textrm{D}]_{r} \otimes [\textrm{II}]_{r}} [2R_{1}^{n}-2R_{2}^{n}]^{2}_{(22)} \ar[d] \ \ \ \ \ \ \ \ \  & 
^{[\textrm{II}]_{r} \otimes [\textrm{D}]_{r}} [2R_{1}^{n}-2R_{2}^{n}]^{2}_{(22)} \ar[d] \\
^{[-] \otimes [\textrm{N}]_{r}} [4R^{n}]^{2}_{(2)} \ \ \ \ \ \ \ \ \  & 
^{[\textrm{N}]_{r} \otimes [-]} [4R^{n}]^{2}_{(2)}
}
\end{displaymath}
\caption{Degeneration scheme of the Type $\textrm{III}_{n}$.}
\end{Scheme}

\subsection{Type IV (1 null eigenvector)}

Finally, for the Type $\textrm{IV}$ we find two subtypes with canonical forms given by
\begin{equation}
C_{ab} = e_{1a}e_{1b} + e_{2a}e_{4b} + e_{4a}e_{2b}
\end{equation}
or
\begin{equation}
C_{ab}  = e_{2a}e_{2b} + e_{1a}e_{4b} + e_{4a}e_{1b}
\end{equation}
The eigenvectors and corresponding eigenvalues are
\begin{displaymath}
e_{4} \longleftrightarrow R^{n} \ , \ \ \ R^{n} =0
\end{displaymath}
The Plebański spinors for the first subtype read
\begin{eqnarray}
V_{ABCD} &=& -2 \, k_{A}k_{B}k_{C}k_{D}
\\ \nonumber
V_{\dot{A}\dot{B}\dot{C}\dot{D}} &=& -4 \, k_{(\dot{A}} k_{\dot{B}}  k_{\dot{C}} l_{\dot{D})}
\end{eqnarray}
Petrov-Penrose types of these spinors are $[\textrm{N}]_{r}$ and $[\textrm{III}]_{r}$, respectively. For the second subtype we find
\begin{eqnarray}
V_{ABCD} &=& -4 \, k_{(A}k_{B}k_{C}l_{D)}
\\ \nonumber
V_{\dot{A}\dot{B}\dot{C}\dot{D}} &=& -2 \, k_{\dot{A}} k_{\dot{B}}  k_{\dot{C}} k_{\dot{D}}
\end{eqnarray}
and the Petrov-Penrose types of Plebański spinors are $[\textrm{III}]_{r}$ and $[\textrm{N}]_{r}$.
\begin{table}[!h]
\begin{center}
\begin{tabular}{|c|c|c|c|}   \hline
\multicolumn{2}{|c|}{PP classification}  & \multicolumn{2}{|c|}{Neutral signature classification}  \\  \hline
Eigenvalues & Type IV & Eigenvalues & Type IV \\ \hline
  $Z^4$ & $[4N]^{a}_{4}$ & $R^4$ & $ ^{[\textrm{N}]_{r} \otimes [\textrm{III}]_{r}} [4R^{n}]^{1}_{(4)}$
  \\ \cline{2-2}  \cline{4-4} 
  & $[4N]^{b}_{4}$ &  & $ ^{[\textrm{III}]_{r} \otimes [\textrm{N}]_{r}} [4R^{n}]^{1}_{(4)}$
  \\ \hline
\end{tabular}
\caption{Subtypes of the Type $\textrm{IV}$.}
\end{center}
\end{table}


\section{Concluding remarks}

In this paper we analyzed the algebraic structure of the traceless Ricci tensor in 4-dimensional spaces equipped with the metric of the neutral signature. Detailed considerations brought us to the conclusion that there are 33 essentially different types of $C_{ab}$ in such spaces. Our classification is purely algebraic. The alternate way of classification of traceless Ricci tensor in Lorentzian spaces was given by R. Penrose \cite{Penrose_klasyfikacja}. It is an interesting question how the Penrose approach can be used in our case. We are going to study this problem soon.

In our work \cite{Chudecki_geometria_strun} some of the types of $C_{ab}$ have been related to the existence of so called, \textsl{congruences of the SD null strings}. Another way of further investigations is to find a more detailed classification of the congruences of SD null strings and relate such a classification with the types of $C_{ab}$ presented here. This question will be investigated elsewhere. 

As mentioned in Introduction we hope that our present work fills the gap left by two papers by Plebański and Przanowski \cite{Plebanski_klasyfikacja_matter,Przanowski_classification} in Acta Physica Polonica.
\newline
\newline
\textbf{Acknowledgments}
\newline
Some points of the present paper were presented in July 2016 in Brno at the 13th conference \textsl{Differential Geometry and its Applications}. The author is indebted to Maciej Przanowski for his interest in this work and for help in many crucial matters.


\end{document}